\documentclass[review,number,sort&compress]{elsarticle}

\usepackage{amssymb}
\usepackage{epsfig}
\usepackage{graphicx}
\usepackage{bm}
\usepackage{dcolumn}
\usepackage{physics}
\usepackage{array, multirow}
\usepackage{amsmath}
\usepackage{lineno}

\journal{NIMA}

\begin{document}

\begin{frontmatter}



\title{Improved RF Measurements of SRF Cavity Quality Factors}


\author{J.P. Holzbauer}
\ead{jeremiah@fnal.gov}
\author{C. Contreras$^\dagger$}
\author{Y. Pischalnikov}
\author{D. Sergatskov}
\author{W. Schappert}
\cortext[ref1]{Operated by Fermi Research Alliance, LLC under Contract No. De-AC02-07CH11359 with the United States Department of Energy.}
\cortext[ref2]{This published material is partially based upon work supported by the U.S. Department of Energy Office of Science Graduate Student Research (SCGSR) program under contract number DE-SC0014664}
\address{Fermi National Accelerator Laboratory*\\P.O. Box 500, Batavia, IL 60510, USA\corref{ref1}\\
$^\dagger$Facility for Rare Isotope Beams**\\Michigan State University\\East Lansing, MI 48824, USA\corref{ref2}}

\begin{abstract}
SRF cavity quality factors can be accurately measured using RF-power based techniques only when the cavity is very close to critically coupled. This limitation is from systematic errors driven by non-ideal RF components. When the cavity is not close to critically coupled, these systematic effects limit the accuracy of the measurements. The combination of the complex base-band envelopes of the cavity RF signals in combination with a trombone in the circuit allow the relative calibration of the RF signals to be extracted from the data and systematic effects to be characterized and suppressed. The improved calibration allows accurate measurements to be made over a much wider range of couplings. Demonstration of these techniques during testing of a single-spoke resonator with a coupling factor of near 7 will be presented, along with recommendations for application of these techniques. 
\end{abstract}

\begin{keyword}

Superconducting RF

\PACS 85.25.Am \sep 84.40.Dc


\end{keyword}

\end{frontmatter}


\section{Power-based RF Cavity Quality Factor Measurements}

\begin{figure}[!ht]
	\centering\includegraphics[clip=true,trim=0cm 0cm 0cm 0cm,width=120mm]{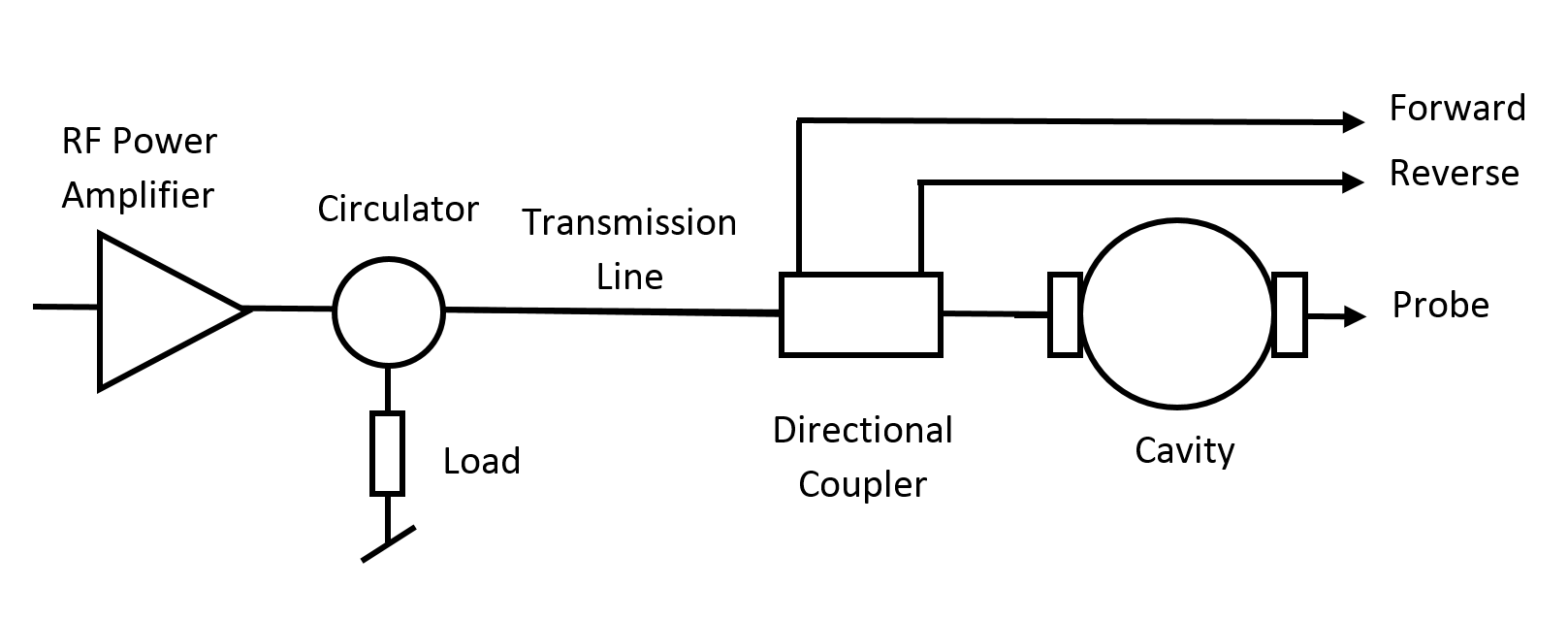}
	\caption{Traditional SRF cavity RF circuit. }
	\label{RFcirc_basic}
\end{figure}

Cavity quality factors are typically measured using a circuit like that shown in Figure \ref{RFcirc_basic} \cite{Powers}. The cavity is driven by an amplifier through a transmission line. A directional coupler separates the forward and reverse waves inside the transmission line.  A field probe monitors the cavity field.

The power of the forward, reverse (sum of waves emitted from the cavity coupler and forward wave reflected from the cavity interface), and transmitted waves during steady state operation on resonance can be combined with measurements of the cavity decay time to determine cavity quality factors if the cavity is close to critically coupled.

The cavity resonant frequency $\omega_0$ and decay time $\tau_{Power}$ can be combined to determine the loaded quality factor, $Q_L$:

\begin{equation}
Q_L = \omega_0 \tau_{Power} = \frac{\omega_0}{2\omega_{1/2}}.
\end{equation}

$Q_L$ is the quality factor of the cavity system which includes fundamental cavity wall losses and power flowing out of the coupler ports and $omega_{1/2}$ is the cavity half bandwidth.  

The reduced cavity coupling factor, $\beta^*$,  can be determined from the cavity reflection coefficient:

\begin{equation}
\beta^* = Q^{-1}_{Ext}\left(Q_0^{-1}+Q_{FP}^{-1}\right)^{-1} = \frac{\sqrt{P^{SS}_{Forward}}\pm\sqrt{P^{SS}_{Reverse}}}{\sqrt{P^{SS}_{Forward}}\mp\sqrt{P^{SS}_{Reverse}}}
\end{equation}

where $SS$ represents Steady State, $Q_0$ is the cavity intrinsic quality factor, $Q_{FP}$ is the field probe coupling quality factor, and $Q_{Ext}$ is the input coupler quality factor. The upper signs are for an over-coupled cavity, and the lower signs are for an under-coupled cavity. 

The cavity field probe coupling can then be determined from $Q_L$, $\beta^*$, and the steady state forward and field probe (FP) power measurements as follows:

\begin{equation}
Q_{FP} = \frac{\omega U}{P_{FP}} = \frac{4Q_L}{1+\beta^{*-1}} \frac{P^{SS}_{Forward}}{P^{SS}_{FP}}
\end{equation}

where $U$ is the stored energy in the cavity. Finally, the intrinsic quality factor, $Q_0$, can be calculated from the above quantities as follows:

\begin{equation}
Q_0 = \frac{Q_{FP} Q_L \left(1+\beta^* \right)}{Q_{FP}-Q_L\left(1+\beta^* \right)}.
\end{equation}

The cavity gradient can be determined from the cavity field probe coupling, field probe (FP) power, cavity impedance $(R/Q)$, and effective length, $L_{Eff}$:

\begin{equation}
E_{Acc} = \frac{\sqrt{Q_{FP} P_{FP}\left(R/Q \right)}}{L_{Eff}}.
\end{equation}

Several systematic effects can bias such measurements \cite{SysErrors}:

\begin{enumerate}
	\item Impedance mismatches between the circulator and the transmission line can reflect part of the reverse wave back into the forward wave. During the decay this can lead to a non-zero forward wave which can interfere constructively or destructively with the cavity field, lengthening or shorting the cavity decay time.
	\item Cross-talk (directivity) in the directional coupler used to separate the forward and reverse waves can cross-contaminate the forward and reverse signals leading to systematic biases in the cavity coupling measurement.
\end{enumerate}

\section{Improving RF-based Quality Factor Measurements}

\begin{figure}[!ht]
	\centering\includegraphics[clip=true,trim=0cm 0cm 0cm 0cm,width=120mm]{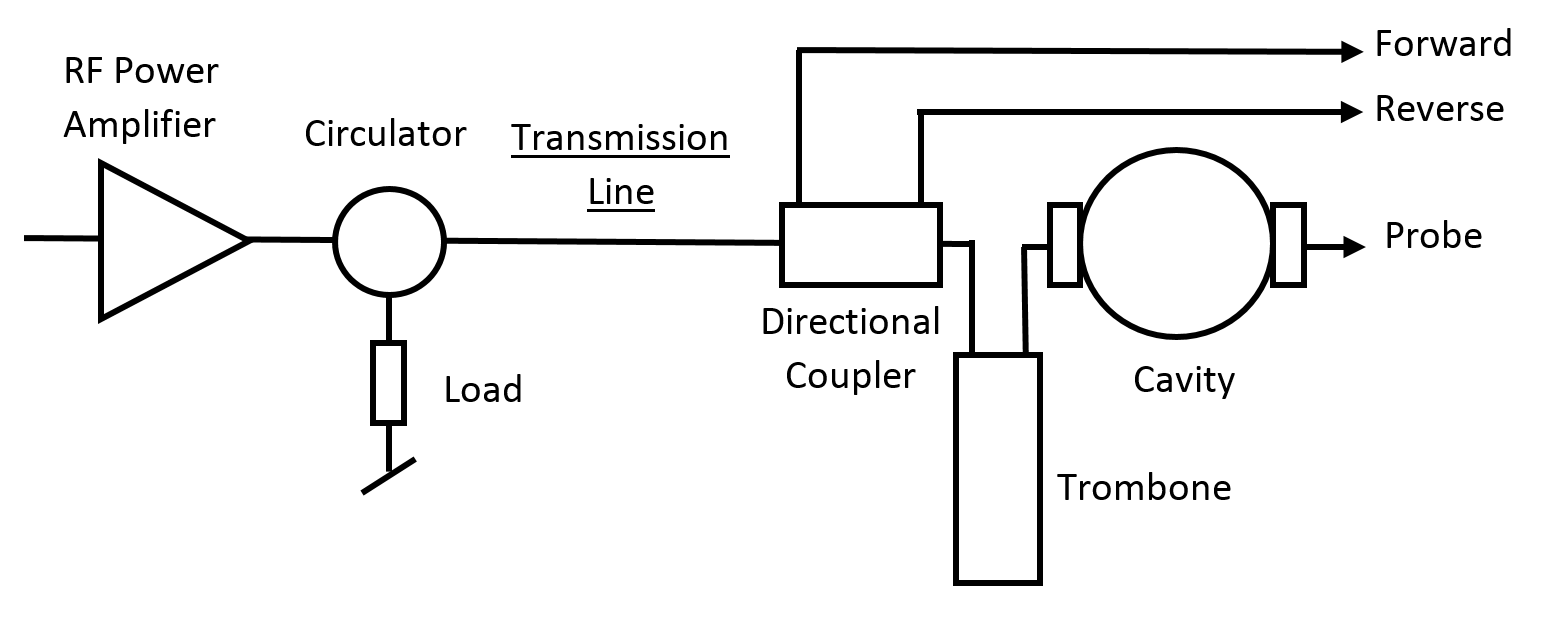}
	\caption{Modified SRF cavity RF circuit including trombone between directional coupler and cavity. This was the RF circuit used during the measurements presented. }
	\label{RFcirc_mod}
\end{figure}

\subsection{Circulator Reflections}

Circulator mismatch causes partial reflection of the reverse wave into the forward wave. Including the reflections from the cavity and circulator, the current in the forward wave is related to the current in the wave emitted by the amplifier as follows \cite{posar}:

\begin{equation}
I_F = \frac{1-\Gamma_{Circulator}}{1-e^{-2i\kappa L}\Gamma_{Circulator}\Gamma_{Cavity}}I_{Amplifier}.
\end{equation}

In this expression, $\Gamma_{Cavity}$ and $\Gamma_{Circulator}$ are respective complex reflection coefficients of the cavity and the circulator, L is the length of the transmission line connecting the circulator and the cavity, and $\kappa = 2\pi/\lambda$ is the RF wavenumber in the transmission line. Given this, $2\kappa L$ represents the round-trip phase advance in the transmission line.  

The true forward wave seen by the cavity can be decomposed into the amplifier wave plus an infinite sum of waves reflected from the cavity/circulator by replacing the numerator with its series expansion around $\Gamma_{Circulator}$.

\begin{equation}
I_F = \sum_{k = 0}^{\infty} I_F^{(k)}
\end{equation}

\begin{equation}
I_F^{(k)} = \left(e^{-2ikL} \Gamma_{Circulator} \Gamma_{Cavity} \right)^k \left(1-\Gamma_{Circulator}\right) I_{Amplifier}
\end{equation}

If the magnitude of the circulator reflection coefficient is much less than one, $|\Gamma_{Circulator}\ll 1$, components other than the first, $I_F^{(1)}$, can be neglected. 

Changing the length of the waveguide by varying the trombone length will change the relative phase of $I_F^{(1)}$ with respect to the direct component, $I_F^{(0)}$. As the trombone length is swept over a wavelength, the relative phase between the direct wave, $I_F^{(0)}$, and the first reflection, $I_F^{(1)}$, will sweep through $4\pi$. The first reflection $I_F^{(1)}$ may interfere constructively or destructively with $I_F^{(0)}$ depending on the phase. This will lead to sinusoidal modulation of both the resonant frequency and decay time of the cavity/waveguide system as the length of the trombone is varied. 

The complex envelope of the cavity field is well described by the differential equation which will be motivated later in Section 3:

\begin{equation}
\frac{dI_P}{dt} = -(\omega_{1/2}-i\delta)I_P+\frac{2\omega_{1/2}}{1+\beta^{-1}}I_F.
\end{equation}

In this expression $\delta = \omega_{Drive}-\omega_0$ is $2\pi$ times the detuning of the cavity, while $I_F$, $I_R$, and $I_P$ represent the respective currents in the forward, reverse, and transmitted (field probe) waves referred to at the plane on the cavity side of the input coupler. In the absence of free charges inside the cavity, Maxwell's equations require that $I_F+I_R+I_P = 0$ at this plane. 

During the decay, the direct, forward wave is turned off, but the circulator reflections persist, continuing to drive the cavity. 
\begin{eqnarray}
I_F|_{Decay} &&= I_F^{(1)}|_{Decay} = e^{-2i\kappa L}\Gamma_{Circulator}I_R|_{Decay}\\&& = I_P|_{Decay} - I_R|_{Decay}\\
&& = \frac{e^{-2i\kappa L}\Gamma_{Circulator}}{1+e^{-2\kappa L}\Gamma_{Circulator}} I_P|_{Decay}
\end{eqnarray}

Substituting this into the envelope equation yields:
\begin{equation}
\frac{d (log I_P|_{Decay})}{dt} = -\omega_{1/2}+i\delta+\frac{2\omega_{1/2}}{1+\beta^{-1}}\frac{e^{-2i\kappa L}\Gamma_{Circulator}}{1+e^{-2i\kappa L\Gamma_{Circulator}}}.
\end{equation}

When $I_F^{(1)}$ is in phase with $I_F^{(0)}$, the measured decay time will be longer than the true decay time of the cavity. Conversely, when $I_F^{(1)}$ is in phase with $I_F^{(0)}$, the measured decay time will shorten:

\begin{equation}
\omega_{1/2}^{Meas} = -\Re{\frac{d(logI_P)}{dt}} = \omega_{1/2}^{Cav}\left(1-\frac{2}{1+\beta^{-1}}\Re{\frac{e^{-2i\kappa L}\Gamma_{Circulator}}{1+e^{-2i\kappa L\Gamma_{Circulator}}}}\right).
\end{equation}

When $I_F^{(1)}$ and $I_F^{(0)}$ are 90 degrees out of phase, the decay time of the cavity/waveguide system will be the same as that of the cavity itself, but the resonant frequency of the system will be shifted with respect to the cavity:

\begin{equation}
\delta^{Meas} = \Im{\frac{d(logI_P)}{dt}} = \delta^{Cav}+\frac{2}{1+\beta^{-1}}\Im{\frac{e^{-2i\kappa L}\Gamma_{Circulator}}{1+e^{-2i\kappa L\Gamma_{Circulator}}}}.
\end{equation}

\subsection{Directional Coupler Directivity}

\begin{figure}[!ht]
	\centering\includegraphics[clip=true,trim=0cm 0cm 0cm 0cm,width=80mm]{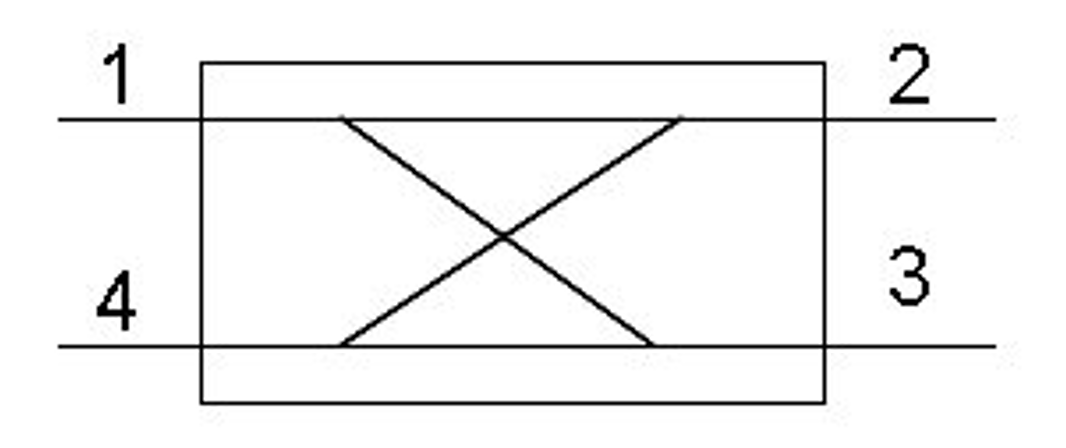}
	\caption{Schematic of a four-port directional coupler. In this model, the waves to be measured travel between ports 1 and 2, with the sampled waves exiting from ports 3 and 4.}
	\label{dircoupler}
\end{figure}

Directional couplers are 4-port linear devices used to separate the forward and reverse waves in a transmission line (Figure \ref{dircoupler}).

Manufacturers typically provide specifications for the insertion loss (IL), isolation (I), coupling (C), and directivity (D):
\begin{eqnarray}
IL &&=10\log(P1/P2) = -20 \log(|S21|)\\
I &&= 10\log(P1/P4) = -20 \log(|S41|)\\
C &&= 10\log(P1/P3) = -20 \log(|S31|)\\
D &&= 10\log(P3/P4) = 20 \log(|S31|/|S41|)
\end{eqnarray}

Directivity is a measure of the cross-talk between the output ports and is defined as the ratio of the powers at the reverse and forward output ports when the device is driven by a wave incident on the forward input port. While directivity provides a useful measure of the magnitude of the cross-talk, a complete understanding of crosstalk in the device also requires knowledge of the phase of the complex S-parameters.

\begin{gather}
\begin{bmatrix} I_{Forward}^{Out} \\ I_{Reverse}^{Out} \end{bmatrix}
=
\begin{bmatrix} S_{31} & S_{32} \\ S_{41} & S_{42} \end{bmatrix}
\begin{bmatrix} I_{Forward}^{In} \\ I_{Reverse}^{In} \end{bmatrix}_.
\end{gather}

These parameters can be determined by changing the length of a trombone inserted between the cavity and the directional coupler. As the length of the trombone is varied, the phase of the forward and reverse waves measured by the directional coupler will change with respect to the forward and reverse waves at the plane of the cavity.

\begin{gather}
\begin{bmatrix} I_{Forward}^{Out} \\ I_{Reverse}^{Out} \end{bmatrix}
=
\begin{bmatrix} S_{31} & S_{32} \\ S_{41} & S_{42} \end{bmatrix}
\begin{bmatrix} e^{i\kappa L_{Trom}} & 0 \\ 0 & e^{-i\kappa L_{Trom}} \end{bmatrix}
\begin{bmatrix} I_{Forward}^{Cavity} \\ I_{Reverse}^{Cavity} \end{bmatrix}_.
\end{gather}

As the length of the trombone is swept over a wavelength, the forward/probe phase will sweep from 0 through 2pi while and reverse/probe phase will sweep from 0 to -2pi.  Any cross-contamination in the measured forward signal will also sweep 0 to -2pi. If the phase and magnitude of the RF signals is recorded during a trombone sweep, the ratios of the complex S parameters, $S_{41}/S_{31}$ and $S_{32}/S_{42}$ can be determined:

\begin{eqnarray}
\frac{S_{41}}{S_{31}} = \frac{\int_{0}^{\frac{2\pi}{\kappa}}dLe^{-i\kappa L}I_{Reverse}^{Out}}{\int_{0}^{\frac{2\pi}{\kappa}}dLe^{-i\kappa L}I_{Forward}^{Out}}\\
\frac{S_{32}}{S_{42}} = \frac{\int_{0}^{\frac{2\pi}{\kappa}}dLe^{-i\kappa L}I_{Forward}^{Out}}{\int_{0}^{\frac{2\pi}{\kappa}}dLe^{-i\kappa L}I_{Reverse}^{Out}}
\end{eqnarray}

When combined with other constraints as discussed below, this procedure allows all elements of the complex directivity matrix,

\begin{gather}
\begin{bmatrix} S_{31} & S_{32} \\ S_{41} & S_{42} \end{bmatrix},
\end{gather}

to be determined up to an overall multiplicative constant. This information can then be used to suppress cross-contamination of the two signals off-line.

\section{Cavity Coupling and Inverse Transfer Functions}

The response of a cavity resonance can be modeled as a RLC oscillator driven from a transmission line at an angular drive frequency, $\omega_{Drive}$, as shown schematically in Figure 5.

\begin{figure}[!ht]
	\centering\includegraphics[clip=true,trim=0cm 0cm 0cm 0cm,width=100mm]{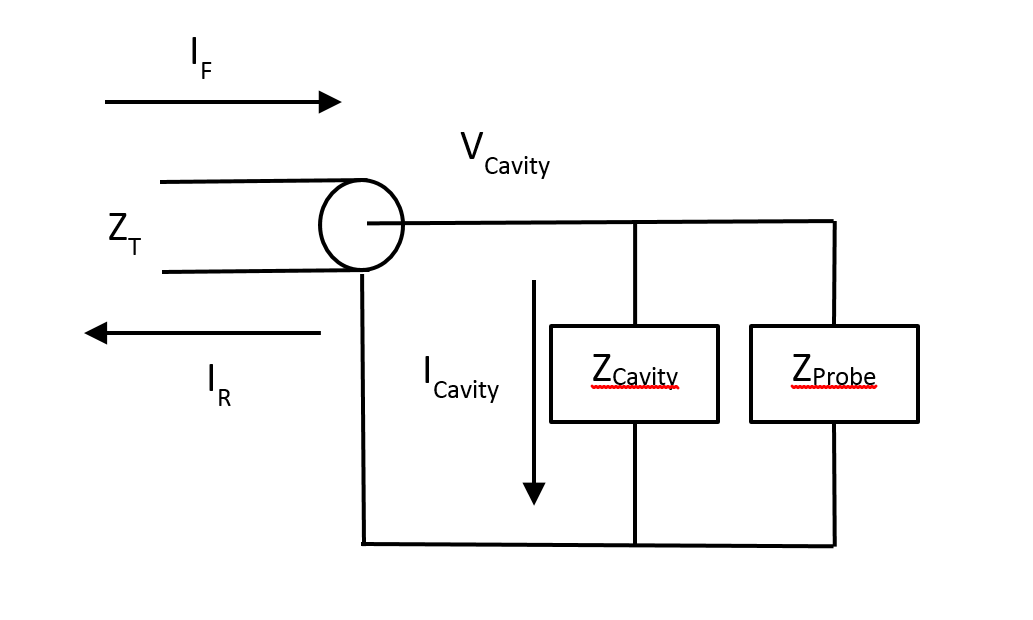}
	\caption{Cavity circuit model including field probe coupler.}
	\label{Cav_circ}
\end{figure}

With the following substitutions:

\begin{equation}
Z_0 = \sqrt{\frac{L}{C}} \equiv \frac{R}{Q}; ~\omega_0 = \frac{1}{\sqrt{LC}}; ~\beta^{-1} = \frac{Z_T}{R_0};
\end{equation}

\begin{equation}
\omega_{Baseband} = \omega' - \omega_{Drive}; ~ \delta = \omega_0 - \omega_{Drive}; ~ \omega_T = \frac{Z_0\omega_{Drive}}{2Z_T} \equiv \frac{\omega_{1/2}}{1+\beta^{-1}};
\end{equation}

the cavity impedance can be written in terms of a dimensionless coupling factor $\beta$,  base-band modulation frequency $\omega_{Baseband}$, detuning $\delta$, and  reduced half bandwidth $\omega_T$:

\begin{equation}
\frac{Z_T}{Z_{Cavity}} = \frac{Z_T}{R_0}+i\frac{Z_T}{Z_0}\left(\frac{\omega}{\omega_0}-\frac{\omega_0}{\omega}\right)
\end{equation}

\begin{equation}
\approx \beta^{-1}+i\frac{\omega_{Baseband}-\delta}{\omega_T}.
\end{equation}

The complex base-band envelopes of the forward and reverse waves at the cavity and the cavity voltage are then related by the following inverse transfer functions (going forward $\omega_{Baseband} = \omega$):

\begin{equation}
T_{P/F}^{-1} = \frac{I_{Forward}^{Cavity}}{Z_T^{-1}V_{Cavity}} = \frac{1}{2}\left(1+\beta^{*-1}+i\frac{\omega-\delta}{\omega_T}\right)
\end{equation}

\begin{equation}
T_{P/R}^{-1} = \frac{I_{Reverse}^{Cavity}}{Z_T^{-1}V_{Cavity}} = \frac{1}{2}\left(1-\beta^{*-1}-i\frac{\omega-\delta}{\omega_T}\right)
\end{equation}

\begin{equation}
\beta^{*-1} = \frac{Z_T}{R_0}+\frac{Z_T}{Z_{FP}}.
\end{equation}

Equation 9 is simply the inverse Fourier transform of Equation 29. 
The sum of the inverse transfer functions adds to unity while the derivatives with respect to detuning are equal and opposite. 

\begin{equation}
T_{P/F}^{-1} + T_{P/R}^{-1} = 1
\end{equation}

\begin{equation}
\frac{\partial T_{P/F}^{-1}}{\partial \delta}+\frac{\partial T_{P/R}^{-1}}{\partial \delta} = 0
\end{equation}

These constraints can be used to determine the relative complex gain of the three cavity RF signals from the data itself.

The real and imaginary components of the difference of the two transfer functions yield the inverse cavity coupling and detuning respectively:

\begin{equation}
T_{P/F}^{-1} - T_{P/R}^{-1} = \beta^{*-1}+i\frac{\omega - \delta}{\omega_T}.
\end{equation}

This allows the inverse coupling to be determined directly from the calibrated inverse transfer functions with no ambiguity whether the cavity is over or under-coupled.

\subsection{Measurement Technique}

The techniques of measuring and correcting for these systematic errors are illustrated using data recorded from a 325 MHz single spoke resonator \cite{SSR1Design, SSR1Testing} installed in the Fermilab STC \cite{STCdesign} operating at 2K with and a nominal gradient of 5 MV/m. The RF circuit was modified by installing a trombone between the directional coupler and the cavity. The length of the trombone was systematically varied over one wavelength (c/325MHz) in 10 steps. At each step of the trombone, the phase of the PLL locking the drive frequency to the cavity resonance frequency was varied in 7 steps between $-45^\circ$ and $45^\circ$ to vary the detuning of the cavity while the complex baseband RF signals were recorded using a digital RF control system provided by the Fermilab AD/LLRF group \cite{STCLLRF} for an interval of 10 seconds. During each recording the generator power was shut off after approximately 7 seconds, allowing the cavity to decay.

For comparison, a control sample was recorded without the trombone in the circuit while the lock phase was varied over the same range.

\section{Measuring and Correcting Directivity}

Cross-contamination of the forward and reverse signals extracted by the directional coupler can lead to significant systematic biases in the determination of the cavity coupling if the cavity is not close to critically coupled.

The measurement of the forward and reverse waves by the directional coupler can be modeled by the product of a trombone dependent phase delay and a linear mixing matrix representing cross-talk (directivity) in the coupler:

\begin{gather}
\begin{bmatrix} I_{Forward}^{DC} \\ I_{Reverse}^{DC} \end{bmatrix}
=
\begin{bmatrix} G_F & \epsilon_F \\ \epsilon_R & G_R \end{bmatrix}
\begin{bmatrix} e^{i\kappa L_{Trom}} & 0 \\ 0 & e^{-i\kappa L_{Trom}} \end{bmatrix}
\begin{bmatrix} I_{Forward}^{Cavity} \\ I_{Reverse}^{Cavity} \end{bmatrix}_.
\end{gather}

The $G$/$\epsilon$ matrix is equivalent to the similar S-Parameter matrix in equation 20. $G_F$ and $G_R$ are the diagonal $S_{31}$ and $S_{42}$ terms and represent the complex gain of the desired directional wave measurement. $\epsilon_F$ and $\epsilon_R$ are the off-diagonal $S_{32}$ and $S_{41}$ terms and represent the complex gain of the cross-contamination of the directional wave measurement. The derivatives of the measured forward/probe signal ratio and measured reverse/probe signal ratio with respect to changes in detuning are related as follows.

\begin{equation}
r_D\left(L_{Trom}\right) = \frac{\frac{\partial}{\partial \delta} \frac{I^{DC}_{Reverse}}{Z^{-1}_T V_{Cavity}}}{\frac{\partial}{\partial \delta} \frac{I^{DC}_{Forward}}{Z^{-1}_T V_{Cavity}}} = -e^{-2i\kappa L_{Trom}}\frac{G_R - \epsilon_R e^{2i\kappa L_{Trom}}}{G_F - \epsilon_F e^{-2i\kappa L_{Trom}}}
\end{equation}

\begin{equation}
\approx \frac{\epsilon_R}{G_F}-\frac{G_R}{G_F}e^{-2i\kappa L_{Trom}}-\frac{G_R}{G_F}\frac{\epsilon_F}{G_F}e^{-4i\kappa L_{Trom}}
\end{equation}

The relative complex gain of the forward and reverse waves and cross-contamination coefficients can be determined by Fourier transforming this ratio with respect to the length of the trombone:

\begin{equation}
R_D(n) = -\sqrt{\frac{\kappa}{2\pi}}\int_{0}^{\frac{2\pi}{\kappa}} dL_{Trom}e^{2i\kappa nL_{Trom}}r_D(L_{Trom}).
\end{equation}

The complex elements of the mixing matrix are then determined up to a single overall complex gain factor, $G_F^{-1}$:

\begin{equation}
\frac{\epsilon_R}{G_F} = R_D(0); ~\frac{G_R}{G_F} = -R_D(-2); ~\frac{\epsilon_F}{G_F} = \frac{R_D(-4)}{R_D(-2)}.
\end{equation}

The overall gain factor can be determined by requiring the inverse transfer functions sum to unity.

\begin{gather}
\begin{bmatrix} T_{P/F}^{-1} \\ T_{P/R}^{-1} \end{bmatrix}
=
G_F^{-1}
\begin{bmatrix} e^{i\kappa L_{Trom}} & 0 \\ 0 & e^{-i\kappa L_{Trom}} \end{bmatrix}
\begin{bmatrix} 1 & \frac{\epsilon_F}{G_F} \\ \frac{\epsilon_R}{G_F} & \frac{G_R}{G_F} \end{bmatrix}^{-1}
\begin{bmatrix} \frac{I^{DC}_{Forward}}{Z^{-1}_T V_{Cavity}} \\ \frac{I^{DC}_{Reverse}}{Z^{-1}_T V_{Cavity}} \end{bmatrix};
\end{gather}
\begin{equation}
T_{P/F}^{-1}+T_{P/R}^{-1} = 1.
\end{equation}

\begin{figure}[!ht]
	\centering\includegraphics[clip=true,trim=0cm 0cm 0cm 0cm,width=120mm]{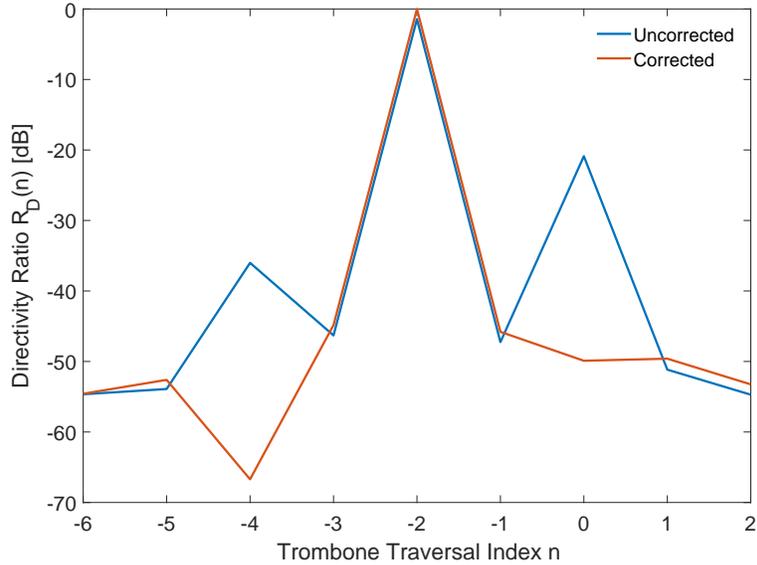}
	\caption{Directivity determination before and after correction factor applied. }
	\label{Dir_spect}
\end{figure}

Equation 39 is not linear in coefficients and thus the procedure must be iterated several times to find a convergent solution of the original rational relationship.

Figure \ref{Dir_spect} compares the transformed derivative ratio measured for the 325 MHz cavity under test before and after directivity determination and correction. As would be expected from the equations, the uncorrected ratio shows a large peak at -2 corresponding to the coefficient ratio $G_R/G_F$  and smaller peaks at -4 and 0 corresponding to the coefficient ratios $\epsilon_F/G_F$  and $\epsilon_R/G_F$. Before correction the directivity is 21 dB.  Following offline correction, the directivity improves to 50 dB giving much better separation and relative calibration of the forward and reverse waves.

\section{Cavity Decay Measurements}

\begin{figure}[!ht]
	\centering\includegraphics[clip=true,trim=0cm 0cm 0cm 0cm,width=120mm]{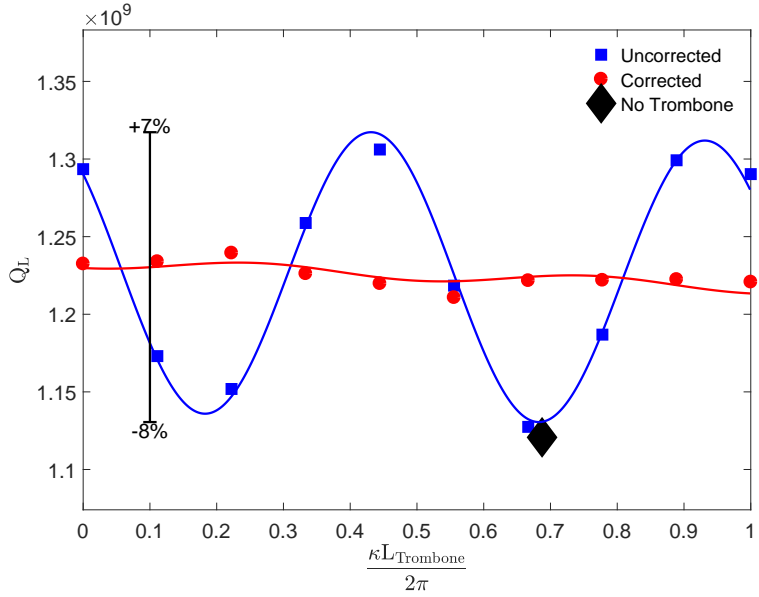}
	\caption{Variation of Loaded Quality Factor $Q_L$ with trombone length (Blue Squares). The $Q_L$ measured without the trombone in the circuit (Black Diamond) agrees with these measurements. Using directivity-corrected powers, the measured $Q_L$ can be corrected (Red Circles) to give results mostly independent of trombone position. }
	\label{QLvsTrom}
\end{figure}

\begin{figure}[!ht]
	\centering\includegraphics[clip=true,trim=0cm 0cm 0cm 0cm,width=120mm]{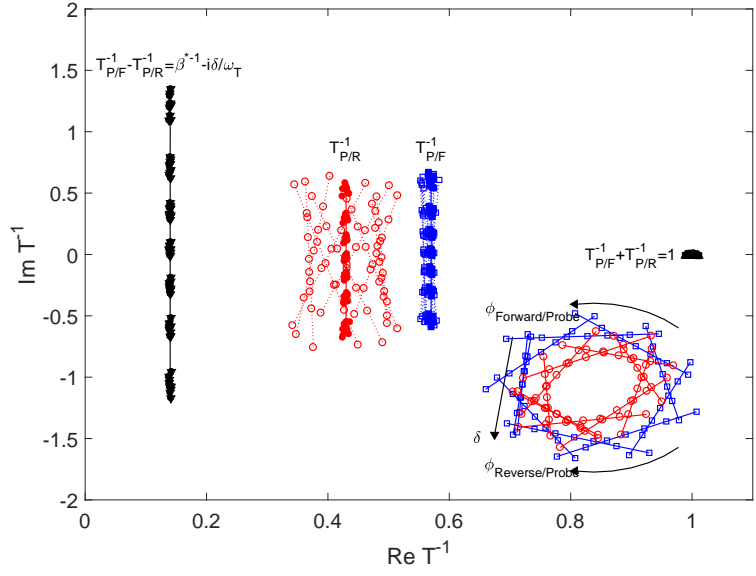}
	\caption{Measured Inverse Transfer Functions (F/P and R/P) as well as their sums and differences. Blue squares represent F/P, red circles are R/P, and black triangles are the sum and differences. Closed symbols are using directivity-corrected signals, and open symbols are uncorrected. The inset data in the lower right shows this data before calibration/directivity correction, showing variation in detuning and trombone position.}
	\label{TFs}
\end{figure}

The blue and red points in Figure \ref{QLvsTrom} respectively compare the loaded cavity quality factor, $Q_L$, as a function of trombone position before and after correction for impedance mismatches at the circulator. The uncorrected measurements were multiplied by the following factor to account for non-zero forward power during the decay to obtain the corrected measurements. This is only possible when the directivity-corrected forward wave is known. Residual, uncorrected components may be due to directivity correction uncertainties, higher order reflections, length-dependent attenuation in the trombone, and measurement noise.

\begin{eqnarray}
\frac{Q_L^{Corrected}}{Q_L^{Uncorrected}} &&= 1+\frac{Re\left\langle \frac{I_F}{I_P}\right\rangle_{Decay}}{Re\left\langle \frac{I_F}{I_P}\right\rangle_{SteadyState}}\\
&& = 1+\frac{2}{1+\beta^{-1}}\frac{e^{-2i\kappa L}\Gamma_{Circulator}}{1+e^{-2\kappa L}\Gamma_{Circulator}}
\end{eqnarray}

The uncorrected measurements (blue squares) vary sinusoidally by up to 8\% over two cycles as the length of the trombone is varied over a wavelength while the corrected measurements (red circles) are much less sensitive to trombone length. 

The black diamond shows the equivalent measurement with no trombone in the circuit. The value is consistent with the value expected from the trombone measurements when the phase length of the waveguide is the same.

\section{Cavity Quality Factor Measurements}

Following correction for directivity and calibration (using the constraints discussed earlier), the cavity inverse transfer functions can be determined from the complex steady state ratios of the forward/probe and reverse/probe signals.

The inverse transfer functions measured for the 325MHz cavity under test operating at 2K are plotted in the complex plane in Figure \ref{TFs}, with and without directivity correction.

\begin{figure}[!ht]
	\centering\includegraphics[clip=true,trim=0cm 0cm 0cm 0cm,width=120mm]{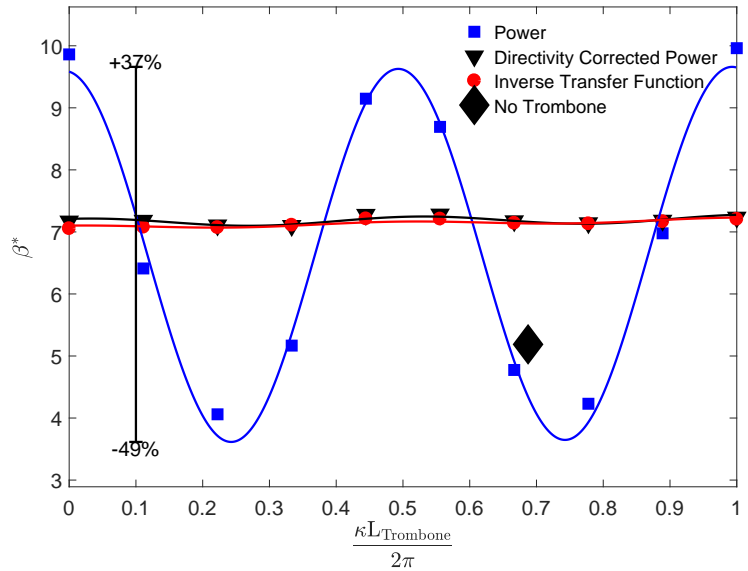}
	\caption{Measured $\beta^*$ via Power measurements versus trombone angle with uncorrected signals (Blue Squares) and directivity-corrected signals (Black Triangles) compared to directivity-corrected Transfer Functions (Red Circles). $\beta^*$ measured without trombone (Black Diamond) agrees with trombone scan data. }
	\label{beta_star}
\end{figure}

The multiple vertical lines for each inverse transfer function represent independent measurements of the two inverse transfer functions as the lock phase is varied between $-45^\circ$ and $45^\circ$ at the each of the different trombone lengths. As the lock is varied the detuning of the cavity changes and the transfer functions sweep along a vertical line in the complex plane. As expected both inverse transfer functions show little sensitivity to changes in the trombone length.

The inverse cavity coupling can be determined directly from the real component of the difference between the two inverse transfer functions:

\begin{equation}
\Re{T_{P/F}^{-1} - T_{P/R}^{-1}} = \beta^{*-1}.
\end{equation}

\begin{figure}[!b]
	\centering\includegraphics[clip=true,trim=0cm 0cm 0cm 0cm,width=120mm]{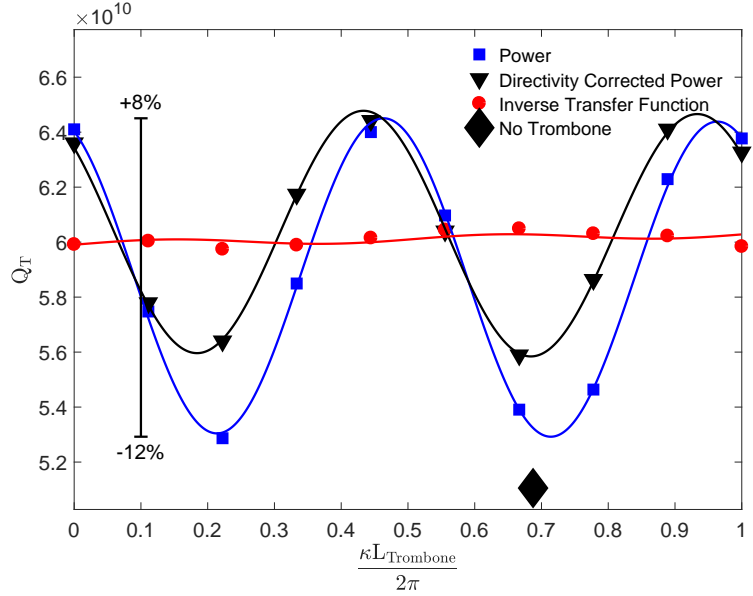}
	\caption{Measured $Q_{FP}$ via Power measurements versus trombone angle with uncorrected signals (Blue Squares) and directivity-corrected signals (Black Triangles) compared to directivity-corrected Transfer Functions (Red Circles). $Q_{FP}$ measured without trombone (Black Diamond) agrees with trombone scan data. }
	\label{QT}
\end{figure}

\begin{figure}[!htb]
	\centering\includegraphics[clip=true,trim=0cm 0cm 0cm 0cm,width=120mm]{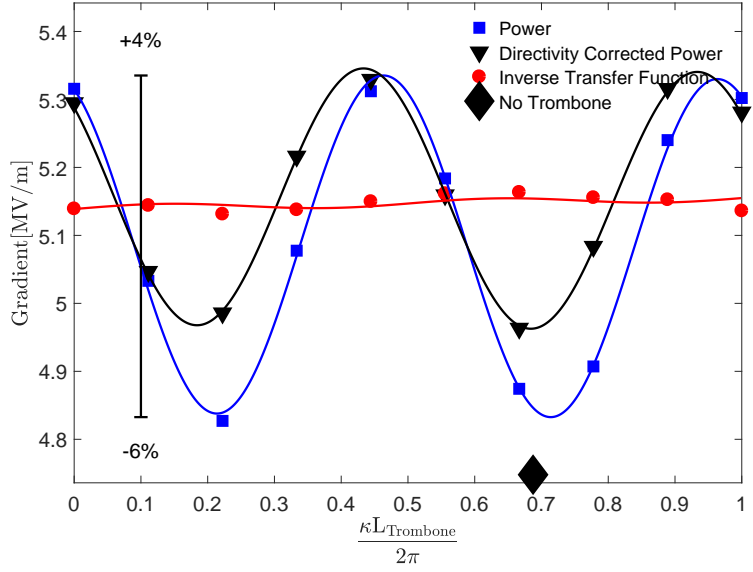}
	\caption{Measured Gradient via Power measurements versus trombone angle with uncorrected signals (Blue Squares) and directivity-corrected signals (Black Triangles) compared to directivity-corrected Transfer Functions (Red Circles). Gradient measured without trombone (Black Diamond) agrees with trombone scan data. }
	\label{Gradient}
\end{figure}

The blue and black points in Figure \ref{beta_star} respectively compare the cavity coupling, $\beta^*$, as a function of trombone position both before and after the cavity power measurements were corrected for directivity. The red points show the inverse cavity coupling determined using the inverse transfer functions.

The blue and black points in Figure \ref{QT} respectively compare the cavity field probe coupling, $Q_{FP}$, as a function of trombone position both before and after the cavity power measurements were corrected for directivity. The red points show the inverse cavity coupling determined using the inverse transfer functions. 

\begin{figure}[!htb]
	\centering\includegraphics[clip=true,trim=0cm 0cm 0cm 0cm,width=120mm]{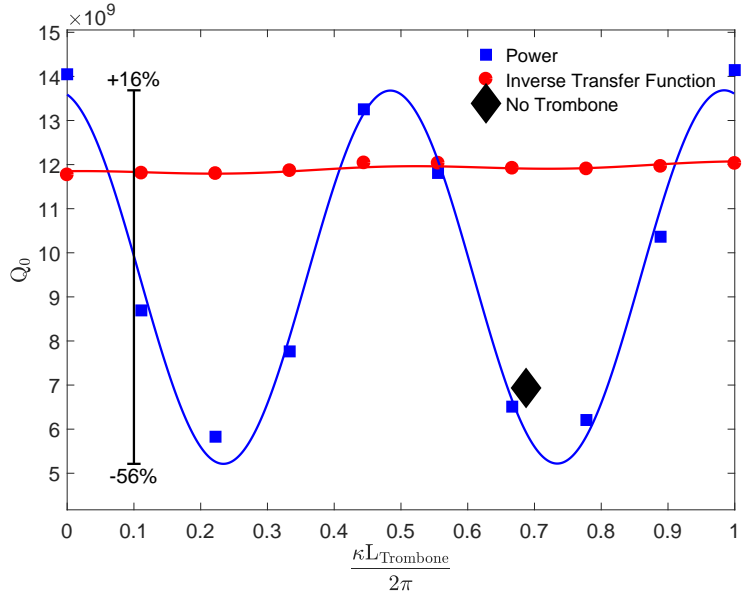}
	\caption{Measured $Q0$ via Power measurements versus trombone angle with uncorrected signals (Blue Squares) compared to directivity-corrected Transfer Functions (Red Circles). $Q_0$ measured without trombone (Black Diamond) agrees with trombone scan data. }
	\label{Q0}
\end{figure}

The uncorrected power measurements (blue) vary sinusoidally by up to 12\% as the length of the trombone is varied over a wavelength while the variation of directivity corrected power measurements (black) is smaller but still pronounced. This is because $Q_{FP}$ is sensitive to both directivity and circulator reflections. The inverse transfer function measurements (red) are much less sensitive to trombone length.  

The black diamond in Figure \ref{QT} again shows the equivalent measurement without a trombone installed in the circuit. Again, the value is consistent with the value expected from the trombone measurements when the phase length of the waveguide is the same.

The blue and red points in Figure \ref{Gradient} respectively compare the cavity gradient, $E_{Acc}$, as a function as a function of trombone position both before and after the cavity power measurements were corrected for directivity. The red points show the inverse cavity coupling determined using the inverse transfer functions. 

The uncorrected power measurements (blue) vary sinusoidally by up to 12\% as the length of the trombone is varied over a wavelength while the variation of directivity corrected power measurements (black) is smaller but still pronounced. The inverse transfer function measurements (red) are much less sensitive to trombone length.  

\begin{figure}[!b]
	\centering\includegraphics[clip=true,trim=0cm 0cm 0cm 0cm,width=120mm]{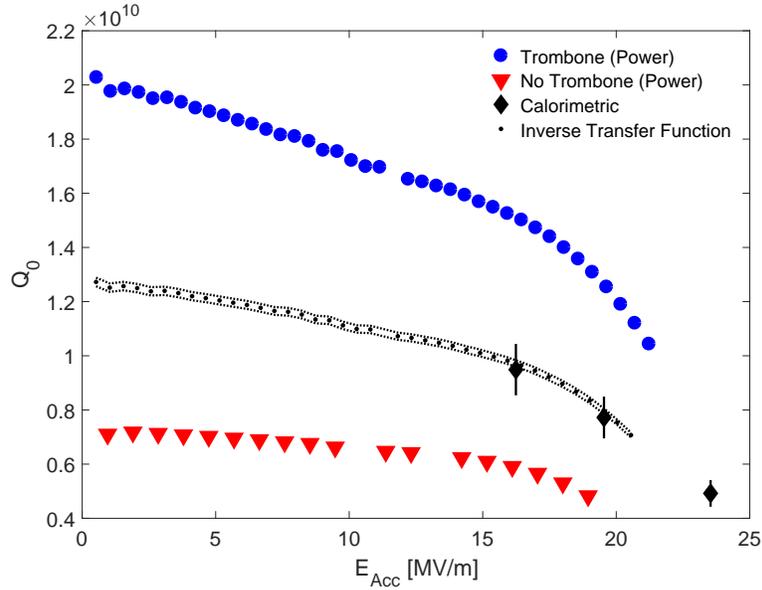}
	\caption{$Q_0$ versus gradient as measured by uncorrected power with no trombone which coinsides with 'minimum $Q_0$ (Red Triangles), at the trombone position which coincides with 'maximum $Q_0$' (Blue Circles), and as measured with directivity-corrected signals via Transfer Functions (Black Dots) with dotted lines represent the error bounds. Calorimetric $Q_0$ measurements (Black Diamonds) agree well with measured $Q_0$.  }
	\label{QvE}
\end{figure}

The black diamond in Figure \ref{Gradient} again shows the equivalent measurement without a trombone installed in the circuit. Again, the value is consistent with the value expected from the trombone measurements when the phase length of the waveguide is the same.

Figure \ref{Q0} compares the intrinsic cavity quality factor, $Q_0$, as determined from power measurements (blue curve) to the same quantity determined using the inverse transfer functions (red curve).

The power measurements vary sinusoidally by up to 56\% as the length of the trombone is varied over a wavelength while the inverse transfer function measurements (red) are much less sensitive to trombone length. 

The black diamond in Figure \ref{Q0} again shows the equivalent measurement with no trombone in the circuit. Again, the value is consistent with the value expected from the trombone measurements at the equivalent phase length.

Figure \ref{QvE} compares the variation of the intrinsic quality factor with accelerating gradient measured using the power technique and the inverse transfer function technique. The two power curves shown (red and blue) represent the extreme limits of the variation of the systematic errors in this test based on trombone position. The coupling calculated from the power in the cavity RF signals depends strongly on the phase length of the waveguide and cross-talk in the directional coupler, varying from 10 to 4 (blue/red curves) at low field. In comparison, the cavity coupling determined from the inverse transfer functions is much less sensitive to both the phase length of the transmission line and coupler directivity.

The cavity quality factor was also measured calorimetrically (black diamonds), using the outlet mass flow meter in the STC cryogenic circuit and in cryostat heaters. Heaters are fired with RF off to calibrate mass flow, then the cavity is held stable at desired gradients. Comparing the mass flow for heaters and RF allows calculation of the cavity quality factor. These measurements are limited by noise in the cryogenic system to above 1-2 Watts, so only high field points were taken. These data points agree very well with the directivity-corrected inverse transfer-function data.  

\subsection{Quality Factor Error Estimation}
The dominant sources of systematic uncertainty in these measurements is believed to arise from:
\begin{enumerate}
	\item Imperfect directivity corrections
	\item Length dependent attenuation in the trombone
	\item Impedance mismatches in the waveguide connecting the trombone to the cavity.
\end{enumerate}

A fit to the corrected $Q_L$ curve in Figure \ref{QLvsTrom} yields fractional uncertainty in $Q_L$ of $\Delta Q_L/Q_L=0.003$, a linear dependence on the trombone length of $\Delta Q_L/Q_L = 0.015\pm 0.006$ per wavelength and a residual fractional sinusoidal dependence on the trombone phase angle of $\Delta Q_L/Q_L =0.003\pm .003$. The length dependence is most likely due to changes in the attenuation of the trombone with length while the sinusoidal dependence is most likely due to residual uncertainties in the directivity correction. 

Reflections from any mismatch between the trombone and the cavity will lead to systematic differences between the forward and reverse waves at the plane of the directional coupler and the true forward and reverse waves at the plane of the cavity.

\begin{figure}[!ht]
	\centering\includegraphics[clip=true,trim=0cm 0cm 0cm 0cm,width=90mm]{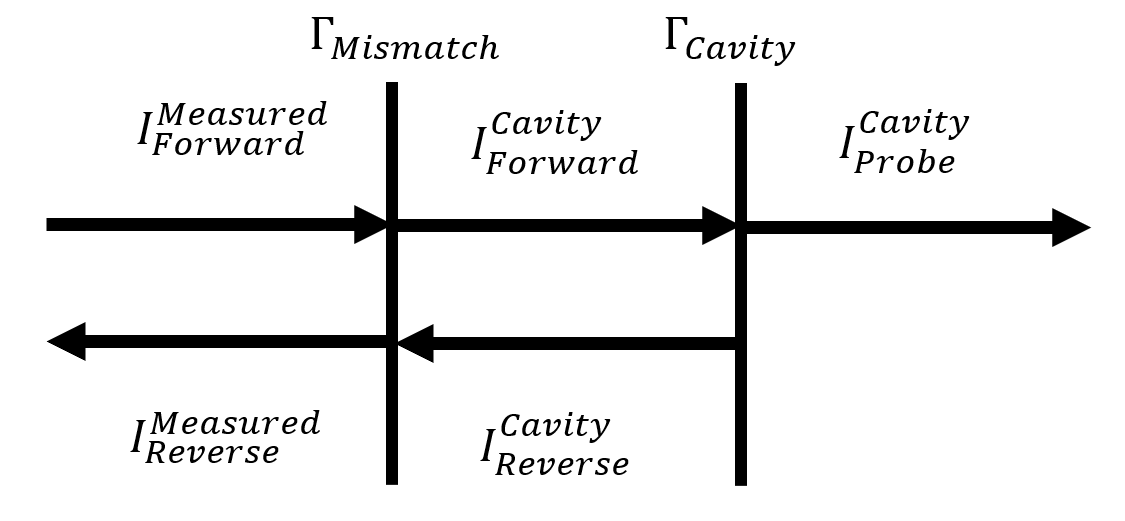}
	\caption{Schematic of the lumped mismatch between the directional coupler and cavity. }
	\label{lump_mismatch}
\end{figure}

Figure \ref{lump_mismatch} shows the circuit model used to estimate the uncertainty associated with these mismatches. For a single mismatch with reflection coefficient, $\Gamma_{Mismatch}$, a distance $L_{Mismatch}$ from the cavity, the waves at the plane of the cavity will be:
\begin{eqnarray}
I_{Forward}^{Cavity} = \left(1-\Gamma_{Cavity}\right)^{-1}I_{Probe}^{Cavity}\\
I_{Reverse}^{Cavity} = \frac{\Gamma_{Cavity}}{\left(1-\Gamma_{Cavity}\right)} I_{Probe}^{Cavity}\\
\Gamma_{Cavity} = \frac{1-\beta^{*-1}-i\frac{\omega-\delta}{\omega_T}}{1+\beta^{*-1}+i\frac{\omega-\delta}{\omega_T}}.
\end{eqnarray}

The forward and reverse waves measured by the directional coupler followed phase correction for the trombone will be:

\begin{eqnarray}
I_{For}^{Meas} = \left(\frac{1-\Gamma_{Mis}}{1-e^{-2i\kappa L_{Mis}}\Gamma_{Mis}\Gamma_{Cav}}\right)^{-1} I_{For}^{Cav}\\
I_{Rev}^{Meas} = \left( 1-\Gamma_{Mis}+ \left(\frac{1-\Gamma_{Mis}}{1-e^{-2i\kappa L_{Mis}} \Gamma_{Mis}\Gamma_{Cav}}\right)^{-1} \frac{\Gamma_{Mis}}{\Gamma_{Cav}} \right) I_{Rev}^{Cav}
\end{eqnarray}
where the super/subscripts have been shortened for space. Keeping only the leading terms in $\Gamma_{Mismatch}$:

\begin{eqnarray}
\frac{I_{For}^{Meas}}{I_{Probe}^{Cav}} = \left(1+\Gamma_{Mismatch} -e^{-2i\kappa L_{Mis}}\Gamma_{Mis}\Gamma_{Cav}\right)\frac{I_{For}^{Cav}}{I_{Probe}^{Cav}}\\
\frac{I_{Rev}^{Meas}}{I_{Probe}^{Cav}} = \frac{I_{Rev}^{Cav}}{I_{Probe}^{Cav}}\\
\beta_{Meas}^{*-1} \approxeq \frac{\Re{\frac{I_{For}^{Meas}}{I_{Probe}^{Cav}}- \frac{I_{Rev}^{Meas}}{I_{Probe}^{Cav}}}}{\Re{\frac{I_{For}^{Meas}}{I_{Probe}^{Cav}}+ \frac{I_{Rev}^{Meas}}{I_{Probe}^{Cav}}}} = \left( 1+ \Gamma_{Mismatch}\left(1-\cos(2\kappa L_{Mis})\right)\right) \beta^{*-1}_{Cav}.
\end{eqnarray} 

The mismatch will also lead to a small residual imaginary component in the sum of the measured inverse forward and reverse transfer functions. 

\begin{equation}
Im\Sigma = \frac{\Im{e^{-i\kappa L_{Mis}}\frac{I_{For}^{Meas}}{I_{Probe}^{Cav}}+ e^{i\kappa L_{Mis}}\frac{I_{Rev}^{Meas}}{I_{Probe}^{Cav}}}}{ \Re{e^{-i\kappa L_{Mis}}\frac{I_{For}^{Meas}}{I_{Probe}^{Cav}}+ e^{i\kappa L_{Mis}}\frac{I_{Rev}^{Meas}}{I_{Probe}^{Cav}}}}\approxeq -\Gamma_{Mismatch}\sin(2\kappa L_{Mis})
\end{equation}

The expected mean square value for the imaginary component of the sum will be:

\begin{equation}
\langle (Im\Sigma)^2\rangle =\frac{|\Gamma_{Mis}|^2}{2} = \frac{2}{3}\left\langle \left( \frac{\Delta\beta^*_{Mis}}{\beta^*}\right)^2\right\rangle.
\end{equation}

The imaginary component of the sum in this dataset was measured to be $Im\Sigma = -0.007$, corresponding to an expected reflection from unknown impedance mismatches of -40 dB and an associated fractional uncertainty in $\beta^* = 0.001$. This scale of mismatch is reasonable given well designed and assembled transmission-line and RF windows. 

The overall expected systematic uncertainty in the measurements can be estimated by combining the estimated systematic errors from all sources in quadrature:
\begin{eqnarray}
\left\langle \left( \frac{\Delta Q_0}{Q_0} \right)^2\right\rangle^{1/2} = \left(\left\langle \left( \frac{\Delta Q_L}{Q_L}\right)^2\right\rangle + \left\langle \left( \frac{\Delta\beta^*}{1+\beta^*} \right)^2 \right\rangle \right)^{1/2}\\
\left\langle \left( \frac{\Delta Q_L}{Q_L}\right)^2\right\rangle^{1/2} = \left\langle \left( \frac{\Delta \tau}{\tau}\right)^2\right\rangle^{1/2}\\
\left\langle \left( \frac{\Delta\beta^*}{1+\beta^*} \right)^2 \right\rangle^{1/2} = \left(\left\langle \left( \frac{\Delta \beta^*_{Dir}}{1+\beta^*}\right)^2\right\rangle + \left\langle \left( \frac{\Delta \beta^*_{Att}}{1+\beta^*}\right)^2\right\rangle + \left\langle \left( \frac{\Delta \beta^*_{Mis}}{1+\beta^*}\right)^2\right\rangle \right)^{1/2}.
\end{eqnarray} 

Estimates of each source of uncertainty are listed in Table \ref{errortable} together with the estimated combined uncertainty. The total uncertainty in this table is the uncertainty represented in Figure \ref{QvE}. 

\begin{table}[]
	\begin{tabular}{lcl}
		\hline
		\multicolumn{1}{|l|}{Source}                                                                        & \multicolumn{1}{l|}{Fractional Uncertainty} & \multicolumn{1}{l|}{Type} \\ \hline
		$\left\langle\left(\frac{\Delta\tau}{\tau}\right)^2\right\rangle^{1/2}$                             & 0.003                            & Statistical               \\
		$\left\langle\left(\frac{\Delta\beta^*_{Dir}}{1+\beta^*}\right)^2\right\rangle$                     & 0.003                            & Statistical               \\
		$\left\langle\left(\frac{\Delta\beta^*_{Att}}{1+\beta^*}\right)^2\right\rangle$                     & 0.006                            & Statistical               \\
		$\left\langle\left(\frac{\Delta\beta^*_{Mis}}{1+\beta^*}\right)^2\right\rangle$                     & 0.010                            & Statistical               \\
		$\left\langle\left(\frac{\Delta Q_0}{Q_0}\right)^2\right\rangle^{1/2}$                               & 0.012                            & Overall                  
	\end{tabular}
\caption{Estimation of the sources of uncertainty in quality factor. }
\label{errortable}
\end{table}

\section{Acknowledgments}

The authors would like to thank Donato Passerelli, Alexander Sukhanov, and Bruce Hanna for helping arrange and execute this test at STC. 

\section{Conclusion}

Systematic effects associated with impedance mismatches at the circulator and imperfect directivity limit the accuracy of cavity quality factor measurements if the cavity coupling is not close to critical. Consistency constraints can be used to improve the calibration of the RF signals if the complex base-band signals are recorded in conjunction with a trombone in the circuit. The improved calibration allows accurate measurements to be made over a wider range of couplings.

The intrinsic quality factor calculated from the cavity power signals from a 325 MHz spoke resonator operating at 2K and a nominal gradient of 5 MV/m with a coupling of $\beta^*$ = 7.14 showed varied between $5\times10^9$ and $1.5\times10^{10}$ depending on the phase length of the transmission line driving the cavity.

When calibrated directivity-corrected complex baseband signals were used to determine the inverse transfer functions and the decay time was corrected for circulator reflections, consistent values for $Q_0$ of $1.19\times10^{10}\pm1.2\%$ at 5 MV/m were obtained independent of the phase length of the waveguide.

\end{document}